\documentclass[aps,pre,preprint,groupedaddress,showpacs]{revtex4-1}
\usepackage[dvips]{graphicx}
\usepackage{textcomp}
\usepackage{amssymb}

\newcommand{\mc}{\multicolumn}

\begin{document}

\title{\Large\bf Spin models in three dimensions: \\
Adaptive lattice spacing}

\author{Martin Hasenbusch}
\email[]{Martin.Hasenbusch@physik.hu-berlin.de}
\affiliation{
Institut f\"ur Physik, Humboldt-Universit\"at zu Berlin,
Newtonstr. 15, 12489 Berlin, Germany}

\date{\today}

\begin{abstract}
Aiming at the study of critical phenomena in the presence of boundaries
with a non-trivial shape we discuss how lattices with an adaptive 
lattice spacing can be implemented.  Since the parameters of the 
Hamiltonian transform non-trivially under changes of the length-scale,
adapting the lattice spacing is much more difficult than in the case of 
the numerical solution of partial differential equations, where 
this method is common practice.

Here we shall focus on the universality class of the three-dimensional 
Ising model. Our starting point is the improved Blume-Capel model
on the simple cubic lattice. In our approach, the system is composed of sectors
with lattice spacing $a$, $2 a$, $4 a$, ... . We 
work out how parts of the lattice with lattice spacing $a$ and $2 a$,
respectively, can be coupled in a consistent way. 
Here, we restrict ourself to the case, where 
the boundary between the sectors is perpendicular to one of the 
lattice-axis.
Based on the theory of defect planes one
expects that it is sufficient to tune the coupling between these 
two regions. To this end we perform a finite size scaling study.
However first numerical results show that slowly decaying 
corrections remain. It turns out that these corrections can be removed 
by adjusting the strength of the couplings within the boundary layers.

As benchmark, we simulate films with strongly symmetry breaking 
boundary conditions. We determine the magnetization profile and 
the thermodynamic Casimir force. For our largest thickness $L_0=64.5$,
we find that results obtained for the homogeneous system are nicely
reproduced.

\end{abstract}

\maketitle

\section{Introduction}
The present work is motivated by the numerical study of effects related to 
critical phenomena in the presence of boundaries with non-trivial shapes. 
Our own interest in this problem stems from an ongoing project on the thermodynamic
Casimir effect \cite{R1,R2,R3,R4} involving finite objects. In particular,
in ref. \cite{mysphere} we studied the thermodynamic Casimir force between
a sphere and a plane by using Monte Carlo simulations of the improved 
Blume-Capel model on a simple cubic lattice. 

In the numerical study of partial differential equations often adaptive
grids are used. In regions, where the field varies strongly, a finer resolution
of the space is used than in regions, where the field is smooth. In fact, 
solving problems related to the thermodynamic Casimir effect 
in the mean-field approximation, a finite element method with a
mesh adapted to the problem is used. For recent work on three-body
interactions see for example refs. \cite{MaHaDi13,MaHaDi14}. 

In the case of critical phenomena, beyond the mean-field approximation, 
this approach is complicated by the 
fact that fields and the strength of the interaction transform in a non-trivial
way under a rescaling of the length. Therefore it is hard to solve the problem in general.
Here we consider the Ising universality class in three dimensions and
in particular study the improved Blume-Capel model.
We have chosen a rather simple setup. We start from a simple cubic lattice with
the lattice spacing $a$. Certain regions of the lattice are coarse grained 
and the lattice spacing is increased to $2 a$, $4 a$, ... . These regions are 
composed of cubes with faces parallel to the $(100)$, $(010)$, and $(001)$ planes.
Here we perform a first step of this program. We study how to couple two half spaces 
with lattice spacing $a$ and $2 a$, respectively. To this end,
we employ finite size scaling to determine the proper values of the couplings
at the boundary between the two different lattice spacings.

In principle the boundary between lattice spacing $a$ and $2 a$ can be seen as a 
defect plane. As argued in refs. \cite{BuEi81,DiDiEi83} such a defect is a relevant 
perturbation of the non-trivial fixed point that characterizes the continuous 
phase transition of the bulk systems. Its RG-exponent is $y_d = y_t -1$, where 
$y_t=1/\nu$ and $\nu$ is the critical exponent of the correlation length of the 
bulk system. This suggests that it is sufficient to tune the coupling between
the regions with lattice spacing $a$ and $2 a$. In our finite-size scaling study
we consider a lattice of the size $L_0 \times L_1 \times L_2$, where
$L_0 = 2 L$ and $L_1=L_2 =L$. We split the lattice in two halfs in $0$-direction.
In one half we use lattice spacing $a$, while in the other one the lattice spacing
is $2 a$.
We define phenomenological couplings that are suitable  to
detect correlations between the two parts of the lattice with different 
lattice spacing.
First numerical results show that there are slowly decaying corrections.
It turns out that these can be eliminated by tuning the couplings
within the lattice planes at the boundary. 

Finally we benchmark our result by simulating films with strongly symmetry breaking 
boundary conditions. We replace half of the sites, located in 
the middle of the film, by coarse grained ones. We compute the thermodynamic Casimir 
force and the magnetization profile. These are compared with estimates that 
we obtained in ref. \cite{myRecent} for the homogeneous system. 

The paper is organized as follows. First we define the Blume-Capel model and 
summarize relevant previous work. 
Next we introduce the finite size scaling method used to determine the 
couplings at the boundary between different lattice spacings. Then we discuss our 
simulations and the numerical results. We summarize our estimates for the couplings
at the boundary between different lattice spacings. Then we present our simulations
of films with strongly symmetry breaking boundary conditions. Finally we conclude
and sketch how we shall use our outcome in upcoming studies of the 
thermodynamic Casimir force.
 
\section{The model}
We consider the Blume-Capel model on a simple cubic lattice. The reduced 
Hamiltonian is given by 
\begin{equation}
\label{Isingaction}
H = -\beta \sum_{<xy>}  s_x s_y
  +  D \sum_x s_x^2 - h \sum_x s_x \;\; ,
\end{equation}
where $s_x \in \{-1, 0, 1 \}$ and $x=(x_0,x_1,x_2)$
denotes a site on the simple cubic lattice, where $x_i$  takes integer values
and $<xy>$ is a pair of nearest neighbors on the lattice. The partition function
is given by $Z = \sum_{\{s\}} \exp(- H)$, where the sum runs over all spin
configurations. The parameter $D$ controls the
density of vacancies $s_x=0$. In the limit $D \rightarrow - \infty$
vacancies are completely suppressed and hence the spin-1/2 Ising
model is recovered. Here we consider a vanishing external field $h=0$ throughout.
In  $d\ge 2$  dimensions the model undergoes a continuous phase transition
for $-\infty \le  D  <  D_{tri} $ at a $\beta_c$ that depends on $ D$.
For $D > D_{tri}$ the model undergoes a first order phase transition, where
for the three-dimensional lattice $D_{tri}=2.0313(4)$ \cite{DeBl04}.

Numerically, using Monte Carlo simulations it has been shown that there
is a point $(D^*,\beta_c(D^*))$
on the line of second order phase transitions, where the amplitude
of leading corrections to scaling vanishes.  Our recent
estimate is $D^*=0.656(20)$ \cite{mycritical}.  In \cite{mycritical} we
simulated the model at $D=0.655$ close to $\beta_c$ on lattices of a
linear size up to $L=360$. From a standard finite size scaling analysis
of phenomenological couplings like the Binder cumulant we find
$\beta_c(0.655)=0.387721735(25)$. Furthermore the amplitude of leading
corrections to scaling is at least by a factor of $30$ smaller than
for the spin-1/2 Ising model. In the following study we shall simulate at 
$D=0.655$ throughout.

\section{The finite size scaling method}
\label{themethod}
We study lattices of the size $L_0=2 L$ and $L_1=L_2=L$ with periodic boundary 
conditions in all three directions. In order to get numbers to compare
with, we simulated a lattice with a unique lattice spacing $a=1$ first.

Next we studied a lattice with the lattice spacing $a=1$ for $0 \le x_0 < L$ and 
lattice spacing $2 a=2$ for $L \le x_0 < 2 L$. 
In the following we shall refer to the fine and the coarse
part of the lattice simply by fine or coarse lattice, respectively.
The sites of the fine lattice are located at 
$x_0=0.5, 1.5, ..., L-0.5$, while those of the coarse lattice assume the 
positions $X_0= L+1, L+3, ..., 2 L-1$.  Between the two parts of the lattice 
there are two boundaries that are perpendicular to the $0$-axis and
are located at $x_0=0$ and $x_0=L$.

\begin{figure}[tp]
\includegraphics[width=13.5cm]{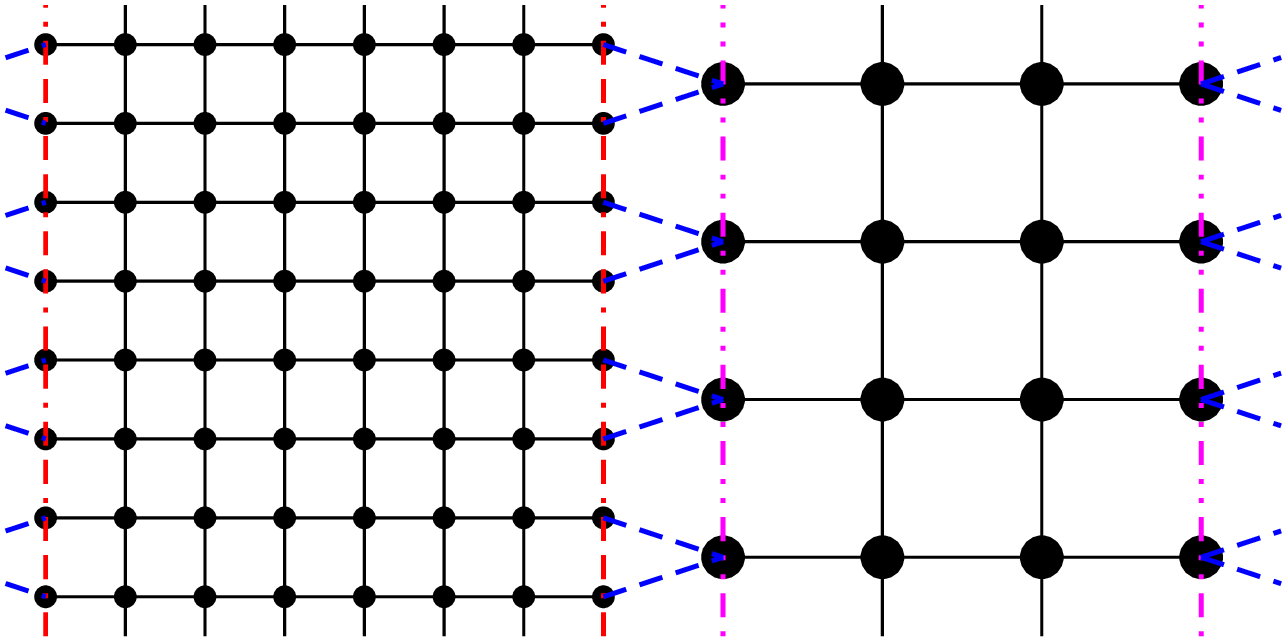}
\caption{\label{sketchlattice}
We give a two-dimensional sketch of the lattice with two lattice spacings.
The couplings in the bulk of the fine and the coarse lattice are denoted by 
$\beta_1$ and $\beta_2$, respectively. The couplings within the layers 
at the boundary between the two parts of the lattice are denoted by $\beta_{1,b}$ and 
$\beta_{2,b}$, respectively. The corresponding links are given by dash-dotted 
lines in the sketch. The coupling between the fine and the coarse lattice 
are denoted by $\beta_{1,2}$. The corresponding links are indicated by 
blue dashed lines.
}
\end{figure}

The coupling constant in the bulk of the fine lattice is called  $\beta_1$, the 
coupling constant in the bulk of the coarse lattice is called $\beta_2$. 
The coupling  between the fine and the coarse lattice is denoted by $\beta_{1,2}$.
It turned out that also the couplings within the layers next to the boundaries
have to be tuned. 
They are denoted by $\beta_{1,b}$ and $\beta_{2,b}$, respectively.
We parameterize these couplings by $\beta_{i,b} =c_{i,b} \beta_{i}$. 
This means that the reduced Hamiltonian of our system is given by
\begin{eqnarray}
\label{mixedaction}
 H &=&  -\beta_1 \sum_{<xy> \in I_1}  s_x s_y
      -\beta_2 \sum_{<XY> \in I_2}  S_X S_Y  
      -\beta_{1,b} \sum_{<xy> \in B_1}  s_x s_y
      -\beta_{2,b} \sum_{<XY> \in B_2}  S_X S_Y \nonumber \\
     && -\beta_{1,2} \sum_{<xY>}  s_x S_Y 
      + D \sum_x s_x^2
      + D \sum_X S_X^2
\end{eqnarray}
where $I_1$ and $I_2$ denote the set of links in the interior of the 
fine and the coarse lattice, respectively.
$B_1$ and $B_2$ denote the set of links within the layers next to the 
boundaries. The subscripts $1$ and $2$ indicate the fine and the coarse
lattice, respectively. Sites of the fine lattice are denoted by
$x$ and $y$, while those of the coarse one are denoted by $X$ and $Y$.
A site in the coarse lattice, adjacent to the boundary, is coupled to 
four sites of the fine one. The spins on the coarse lattice are $S_X \in \{-1,0,1\}$.
In Fig. \ref{sketchlattice} we give a two-dimensional sketch. 

Next we define the quantities that we computed. We  consider the magnetization 
in the center of the fine lattice 
\begin{equation}
 M_1 = \sum_{ L/4 < x_0 < 3 L/4,x_1,x_2} s_x
\end{equation}
and the magnetization in the center of the coarse lattice
\begin{equation}
 M_2 = \sum_{L+L/4 < X_0 < L + 3 L/4,X_1,X_2} s_X  \;\;.
\end{equation}
It is plausible that the normalized correlation function 
\begin{equation}
\label{Gdef}
 G_{1,2} = \frac{\langle M_1 M_2 \rangle}{\sqrt{\langle M_1^2 \rangle \langle M_2^2 \rangle}}
\end{equation}
is well suited to tune the coupling $\beta_{1,2}$ between the two 
lattices. In addition we determined the fourth order cumulant
\begin{equation}
\label{Vdef}
 V_i = \frac{\langle M_i^4 \rangle}{\langle M_i^2 \rangle^2}
\end{equation}
where $i \in \{1,2\}$.

We study the expectation value of the square of the magnetization
\begin{equation}
\label{qdef}
 q(x_0)  =  \frac{1}{L^3} \left \langle \left(\sum_{x_1,x_2} s_x \right)^2 \right \rangle \;\;,
\end{equation}
where the normalization is chosen such that $q(x_0)$ varies slowly with $L$. 
At the critical point we get
\begin{equation}
q(x_0)  \propto L^{-\eta} \;\;.
\end{equation}
Furthermore we consider the profiles of the energy density
\begin{equation}
\label{E0def}
 E_0(x_0)  = \frac{1}{L^2} \sum_{x_1,x_2} \langle s_{x_0,x_1,x_2} s_{x_0+1,x_1,x_2} \rangle
\end{equation}
and
\begin{equation}
\label{E12def}
 E_{12}(x_0)  = \frac{1}{L^2} 
\sum_{x_1,x_2} \langle s_{x_0,x_1,x_2} s_{x_0,x_1+1,x_2} + s_{x_0,x_1,x_2} s_{x_0,x_1,x_2+1}
       \rangle  \;\;.
\end{equation}
Analogous definitions are used for the coarse lattice.  In the case of a homogeneous
system, $q(x_0)$, $E_0(x_0)$, and $E_{12}(x_0)$ do not depend on $x_0$.  In the system with
lattice spacings $a=1$ and $2 a=2$ for the two parts of the lattice, this behavior should be 
reproduced as good as possible. Deviations should rapidly decay with the distance
from the boundaries between different lattice spacings.

\section{Numerical results}
\subsection{Simulations at the critical point}
First we simulated at our best estimate
of the critical coupling $\beta_1=\beta_2=0.387721735$, ref. \cite{mycritical}. 
As a preliminary step we determined the quantities $G_{1,2}$, $V$, $q$, $E_0$, and $E_{12}$
defined above for lattices with a unique lattice spacing $a=1$ for 
$L=8$, $12$, $16$, $24$, ..., and $128$. For equilibration we performed $10^5$ times 
two sweeps with the local heat-bath algorithm followed by $L$ single cluster updates \cite{Wolff}. Performing a sweep means that we run through the lattice in type-writer fashion.
Then we evaluated the quantities of interest after two sweeps
with the local heat-bath algorithm followed by $L$ single cluster updates \cite{Wolff}. 
In the following we shall refer to the evaluation of the quantities as performing 
a measurement. The number of these measurements for each lattice size is given in table \ref{VGLTAB}, where we also summarize our estimates of the 
observables. In total these simulations 
took about 8 years of CPU-time on a single core of a Quad-Core AMD Opteron(tm) 2378 CPU.

\begin{table}
\caption{\sl \label{VGLTAB}
Results obtained for lattices with a unique lattice spacing at $\beta=0.387721735$. 
The value of the energy density in the limit $L \rightarrow \infty$ is taken from 
ref. \cite{myamplitude}. In the first column we give the lattice size $L$.  
In the second column we give the number of measurements divided by $10^6$. In the  following
columns we give our numerical estimates  of the quantities defined in 
eqs. (\ref{Gdef},\ref{Vdef},\ref{qdef},\ref{E0def},\ref{E12def}), respectively.
}
\begin{center}
\begin{tabular}{rrlllll}
\hline
\mc{1}{c}{$L$}  & \mc{1}{c}{stat/$10^6$} & \mc{1}{c}{$G_{1,2}$} & \mc{1}{c}{$V$}  &  \mc{1}{c}{$E_0$} &  
 \mc{1}{c}{$E_{12}$} & \mc{1}{c}{$q$}  \\
\hline
 12  &  1005 & 0.621630(27) &  1.794209(46) &   0.2126616(18) &  0.2129622(18)  & 0.933959(32) \\
 16  &  1015 & 0.623497(27) &  1.794036(47) &   0.2087201(13) &  0.2088462(13)  & 0.926344(33) \\
 24  &  1008 & 0.624844(28) &  1.793880(50) &   0.20524821(85)  &  0.20528524(85)   & 0.914493(36) \\
 32  &   921 & 0.625336(30) &  1.793806(54) &   0.20373756(64)  &  0.20375324(64)   & 0.905711(39) \\
 48  &   816 & 0.625699(34) &  1.793627(60) &   0.20241808(43)  &  0.20242275(43)   & 0.893115(43) \\
 64  &   380 & 0.625721(50) &  1.793701(90) &   0.20184545(45)  &  0.20184739(45)   & 0.883943(65) \\
 96  &   302 & 0.625975(59) &  1.79355(11)  &   0.20134865(31)  &  0.20134924(31)   & 0.871432(76) \\
128  &   270 & 0.626055(64) &  1.79347(12)  &   0.20113357(23)  &  0.20113379(23)   & 0.862552(83)  \\
\hline  
\end{tabular}
\end{center}
\end{table}

Then for systems with two different lattice spacings we tuned the couplings at the 
boundary between the two lattice spacings such that we reproduce the values of the 
quantities $G_{1,2}$ and $V_i$. The quantities $q$, $E_0$ and $E_{12}$
are studied to further validate the results obtained.
Also here we performed a measurement after two sweeps with the heatbath algorithm 
followed by $L$ single cluster-updates. As seed of the cluster we take with
probability $1/3$ a site of the coarse lattice and with probability $2/3$
a site of the fine lattice. This way the average volume of the clusters in the 
fine and the coarse lattice is approximately equal.
In addition to the quantities defined in
eqs. (\ref{Gdef},\ref{Vdef},\ref{qdef},\ref{E0def},\ref{E12def}) we have implemented
the correlators needed to compute the first derivative of these quantities with
respect to $\beta_{1,2}$.  With hindsight we would have also implemented the 
derivatives with respect to the other couplings.  For given $c_{1,b}$ and $c_{2,b}$,
we determined the proper value of $\beta_{1,2}$ by requiring
\begin{equation}
\label{tunebeta12}
 G_{1,2}(L,\bar{\beta}_{1,2}) = \bar{G}(L) := \frac{G_{1,2,hom}(L/2) + G_{1,2,hom}(L)}{2} 
\end{equation}
where the subscript $hom$ indicates the homogeneous system. 
With our definition of $\bar{G}(L)$ we aim at a cancelation of corrections 
to scaling to a large extend. In the following we shall indicate results taken 
from the homogeneous system by a bar. We solved eq.~(\ref{tunebeta12})
iteratively. Starting from a first guess $\bar \beta_{1,2}^{(0)}$ for the solution of
eq.~(\ref{tunebeta12}) we iterated
\begin{equation}
 G_{1,2}(L,\bar \beta_{1,2}^{(i)}) + \frac{\mbox{d} G_{1,2}}{\mbox{d} \beta_{1,2}} 
\left(\bar \beta_{1,2}^{(i+1)} - \bar \beta_{1,2}^{(i)} \right)
= \bar{G}(L)
\end{equation}
until $\bar \beta_{1,2}^{(i+1)} - \bar \beta_{1,2}^{(i)}$ is small. The number of measurements
at a given $\bar \beta_{1,2}^{(i)}$ is increased as we approach the solution.
For larger lattice sizes we took $\bar \beta_{1,2}^{(0)}$ from the 
results that we already obtained for smaller ones.

\begin{figure}[tp]
\vskip1.0cm
\includegraphics[width=13.5cm]{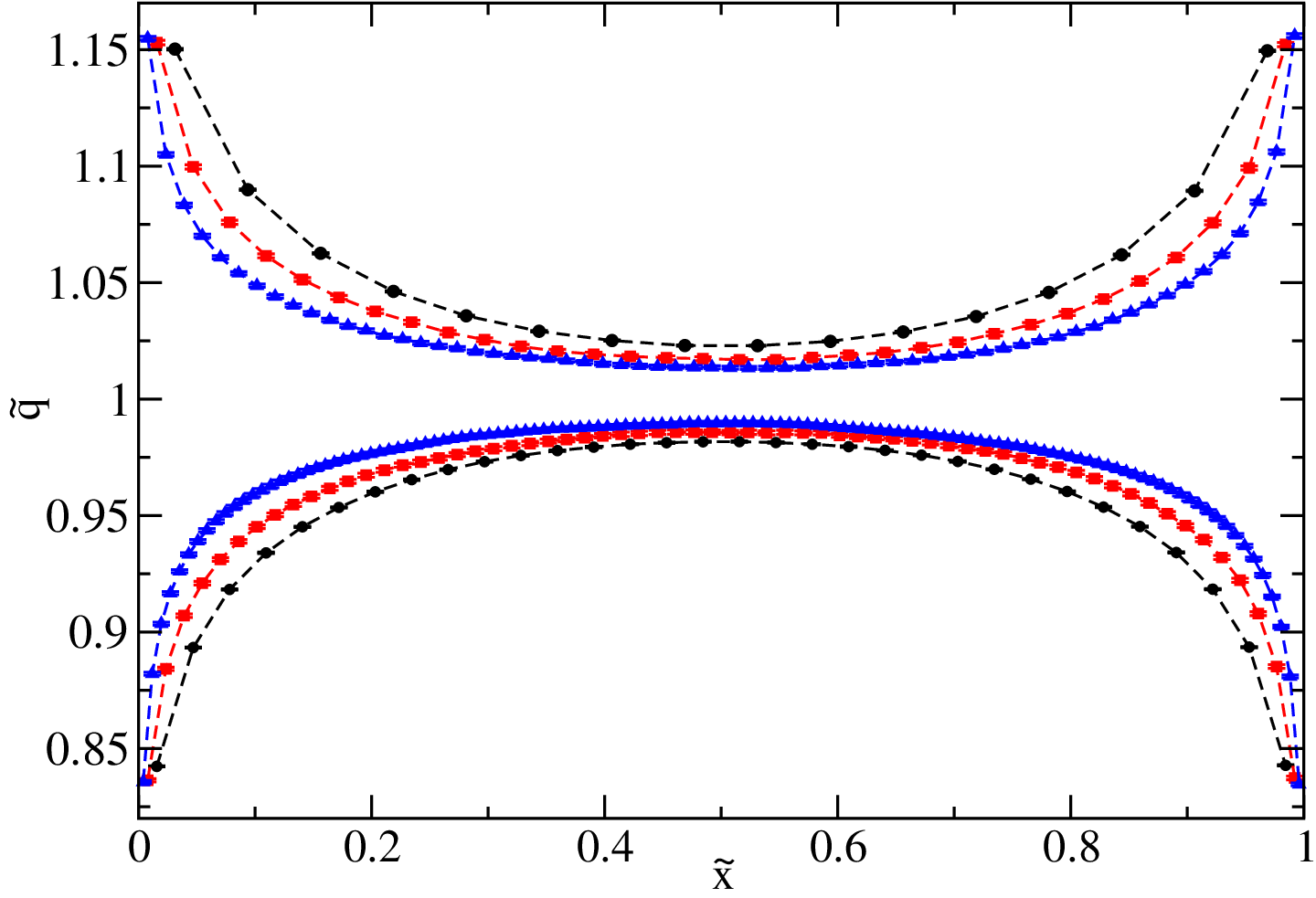}
\caption{\label{magplot}
Profiles, eq.~(\ref{qdef}), at $c_{1,b}=c_{2,b} = 1$ and the lattice sizes
$L=32$, $64$ and $128$.  
We plot $\tilde q=q(x_0)/\bar{q}$ as a function of $\tilde x= x_0/L$ and $ \tilde x= (x_0-L)/L$ for the 
fine and the coarse lattice, respectively. The curves for the different lattice sizes can be
identified by the number of data points. Furthermore $\tilde q < 1$ for the fine lattice
and $\tilde q > 1$ for the coarse one. The dashed lines should only guide the eye.
}
\end{figure}

First we simulated at $c_{1,b}=c_{2,b} = 1$ systems with the lattice sizes $L=32$, $64$ and
$128$. At the resulting values of $\bar \beta_{1,2}$ we computed also the other 
quantities of interest. In Fig. \ref{magplot} we plot $q(x_0)/\bar{q}$,
where $\bar{q}$ is the estimate obtained for the homogeneous system with  
the lattice size $L$ and $L/2$ for the fine and the coarse lattice, respectively.

We see that there are large corrections at the boundaries. For the 
fine lattice $q(x_0)$ is too small and too large for the coarse lattice.
This effect only slowly decreases with increasing lattice size.
This rather strong deviation from the behavior of the homogeneous
system can also be seen clearly in the estimates of $V_1$ and $V_2$
reported in table \ref{V12}.  We also give the estimates of
\begin{equation}
\label{RVdef}
 R_V =\frac{V_1}{V_2} - 1 
\end{equation} 
which allows us to monitor the asymmetry between the fine and the coarse 
lattice.

\begin{table}
\caption{\sl \label{V12}
Results for $\bar \beta_{1,2}$, $V_1$ and  $V_2$ at $c_{1,b}=c_{2,b} = 1$ and $\beta_c$.
In the last column we give $R_V$ as defined in eq.~(\ref{RVdef}).
}
\begin{center}
\begin{tabular}{ccccc}
\hline
$L$  &   $\bar \beta_{1,2}$& $V_1 - \bar{V} $ & $V_2 - \bar{V} $ & $R_V$ \\
\hline
 32 &   0.191000(23) &  0.0213(3)  & --0.0265(3) &0.0262(2) \\ 
 64 &   0.190598(48) &  0.0169(8)  & --0.0194(7) &0.0200(6) \\
128 &   0.190505(25) &  0.0121(7)  & --0.0147(6) &0.0147(5) \\
\hline
\end{tabular}
\end{center}
\end{table}

We see that $|V_1-\bar{V}|$, $|V_2-\bar{V}|$, and $|R_V|$ are decreasing rather slowly with increasing 
lattice size $L$. Fitting with the Ansatz
\begin{equation}
 V_1-\bar{V} = c L^{-\omega_V} 
\end{equation}
we obtain $\omega_V \approx 0.4$. For $V_2-\bar{V}$ and $R_V$ we get similar results.

It turns out that this deviation from the desired behavior can be lifted to a 
large extend by tuning $c_{1,b}$ and $c_{2,b}$. We  employ the quantities 
$V_1$ and $V_2$ to this end. By construction, they are not sensitive to corrections 
that decay rapidly with increasing distance from the boundary between the
fine and the coarse lattice.

First we searched for choices of $c_{1,b}$ and $c_{2,b}$, where $R_V$ vanishes.
Our simulations show that this is the case for a line in the 
$(c_{1,b},c_{2,b})$-plane.
At the level of the precision of our preliminary study this is the case for e.g. 
$(c_{1,b},c_{2,b}) =(1,0.843)$, $(1.072,0.93)$, $(1.128,1)$. However, it turns out that 
in general still $V_1 \approx V_2$ deviate from the value obtained for the homogeneous
system. For $(c_{1,b},c_{2,b})=(1,0.843)$ we get $V_1 -\bar{V} = 0.0052(3)$, $0.0025(4)$, 
and $0.0021(6)$  for $L=$
$32$, $64$, and $128$, respectively. On the other hand, for 
$(c_{1,b},c_{2,b}) = (1.128,1)$ we get  $V_1 -\bar{V} = -0.0059(3)$, $-0.0028(4)$,
and $-0.0019(5)$ for $L=$ $32$, $64$, and $128$, respectively. 
For $(c_{1,b},c_{2,b}) =(1.072,0.93)$ we find that $V_1 -\bar{V}$ approximately vanishes.

Motivated by this preliminary study, 
we systematically searched for the $(c_{1,b},c_{2,b})$, where both
$V_1 - \bar{V}$ and $V_2 - \bar{V}$ vanish  at $\bar \beta_{1,2}$.
To this end we simulated systems with $L=24$, $32$ 
and $48$ for various choices of $(c_{1,b},c_{2,b})$ and interpolated 
$V_1 - \bar{V}$ and $V_2 - \bar{V}$ linearly in $(c_{1,b},c_{2,b})$.
Our results are summarized in table \ref{finalcbtab}. 
\begin{table}
\caption{\sl \label{finalcbtab}
Estimates of the optimal $(c_{1,b},c_{2,b})$ at $\bar \beta_{1,2}$ and $\beta_1=\beta_2=\beta_c$
as a function of the lattice size $L$. We also give the sum and the difference of the optimal
$c_{1,b}$ and $c_{2,b}$. 
}
\begin{center}
\begin{tabular}{cllll}
\hline
$L$  & \mc{1}{c}{$c_{1,b}$}  &  \mc{1}{c}{$c_{2,b}$} & $c_{1,b}+c_{2,b}$ & $c_{1,b}-c_{2,b}$  \\
\hline
24 & 1.0685(5) &0.9237(5) & 1.9922(8) &  0.1448(3)  \\
32 & 1.0679(6) &0.9233(7) & 1.9912(11)&  0.1446(3)  \\
48 & 1.0681(9) &0.9235(12)& 1.9916(20)&  0.1446(4)  \\ 
\hline
\end{tabular}
\end{center}
\end{table}
It turns out that the difference  $c_{1,b}-c_{2,b}$ has a much smaller error than 
the sum $c_{1,b}+c_{2,b}$. This might correspond to the fact that for the line 
where $R_V = 0$ the difference $c_{1,b}-c_{2,b}$ is roughly constant.

Since the estimates obtained for $L=24$, $32$ and $48$ are 
consistent within errors, we abstained from studying larger 
lattice sizes at this point. As our final choice we take 
\begin{equation}
\label{finalcb}
(c_{1,b},c_{2,b}) =(1.068,0.9234) \;.
\end{equation}

We had simulated at $L=24$, $32$ and $48$ at 
$(c_{1,b},c_{2,b}) =(1.068,0.923)$ and $(1.068,0.924)$ with a high 
statistics of $10^8$ measurements or more. In table \ref{finalfinal}
we interpolate the numerical results to $(c_{1,b},c_{2,b}) =(1.068,0.9234)$.
Furthermore we report the results of simulations performed directly at 
$(c_{1,b},c_{2,b}) =(1.068,0.9234)$ for $L=64$, $96$ and $128$. 
For these lattice sizes we performed $94 \times 10^6$, $61 \times 10^6$
and $46 \times 10^6$ measurements, respectively. The simulations 
of the $L=128$ lattice took about 5.4 years of CPU-time
on a single core of a Quad-Core AMD Opteron(tm) 2378 CPU.

\begin{table}
\caption{\sl \label{finalfinal}
Results for $\bar \beta_{1,2}$, $V_1 - \bar{V}$ and $V_2 - \bar{V}$ at 
the optimal $(c_{1,b},c_{2,b})=(1.068,0.9234)$ at $\beta_1=\beta_2=\beta_c$. 
For a discussion see the text.
}
\begin{center}
\begin{tabular}{rcrrr}
\hline
\mc{1}{c}{$L$} &  $\bar \beta_{1,2}$ &\mc{1}{c}{$V_1 - \bar{V}$} &\mc{1}{c}{$V_2 - \bar{V}$} \\
\hline
  24  &  0.1858875(47) &  0.00010(6)\phantom{0}& --0.00000(5)\phantom{0} \\
  32  &  0.1858599(43) &--0.00003(6)\phantom{0}&   0.00000(6)\phantom{0} \\
  48  &  0.1858328(39) &  0.00001(7)\phantom{0}&   0.00001(6)\phantom{0} \\
  64  &  0.1858187(46) &--0.00002(11)&  0.00002(8)\phantom{0} \\
  96  &  0.1858184(45) &--0.00007(13)&  0.00000(9)\phantom{0}  \\
 128  &  0.1858099(45) &--0.00001(15)&  0.00002(13) \\
\hline
\end{tabular}
\end{center}
\end{table}

We find that for $L=64$, $96$ and $128$ the estimates of $V_1 - \bar{V}$  and $V_2 - \bar{V}$ 
remain consistent with zero.
We fitted our data for $\bar \beta_{1,2}$ with the Ansatz
\begin{equation}
\bar \beta_{1,2}(L) = \bar \beta_{1,2}(\infty) + c L^{-\epsilon} \;.
\end{equation}
Taking $\epsilon=2$ and the data for $L=48$, $64$, $96$ and $128$ we arrive at
$\bar \beta_{1,2}(\infty)=0.1858085(39)$, $c=0.054(14)$ and $\chi^2$/d.o.f.$=0.36$. 
In order to check the dependance of the result on $\epsilon$ we repeated the fit
with $\epsilon=1.6$. We arrive at 
$\bar \beta_{1,2}(\infty)=0.1858061(45)$, $c=0.013(3)$ and $\chi^2$/d.o.f.$=0.38$.
Since the estimates of $\bar \beta_{1,2}$ obtained for $L=64$, $96$ and $128$ are
consistent within errors, it is not surprising that $\bar \beta_{1,2}(\infty)$ does not
depend strongly on $\epsilon$. As our final estimate we take
\begin{equation}
\bar \beta_{1,2}(\infty) = 0.185807(6) \;.
\end{equation}
where the error that we quote also takes into account the systematic uncertainty 
of the extrapolation to $L \rightarrow \infty$.

\begin{figure}[tp]
\vskip1.0cm
\includegraphics[width=13.5cm]{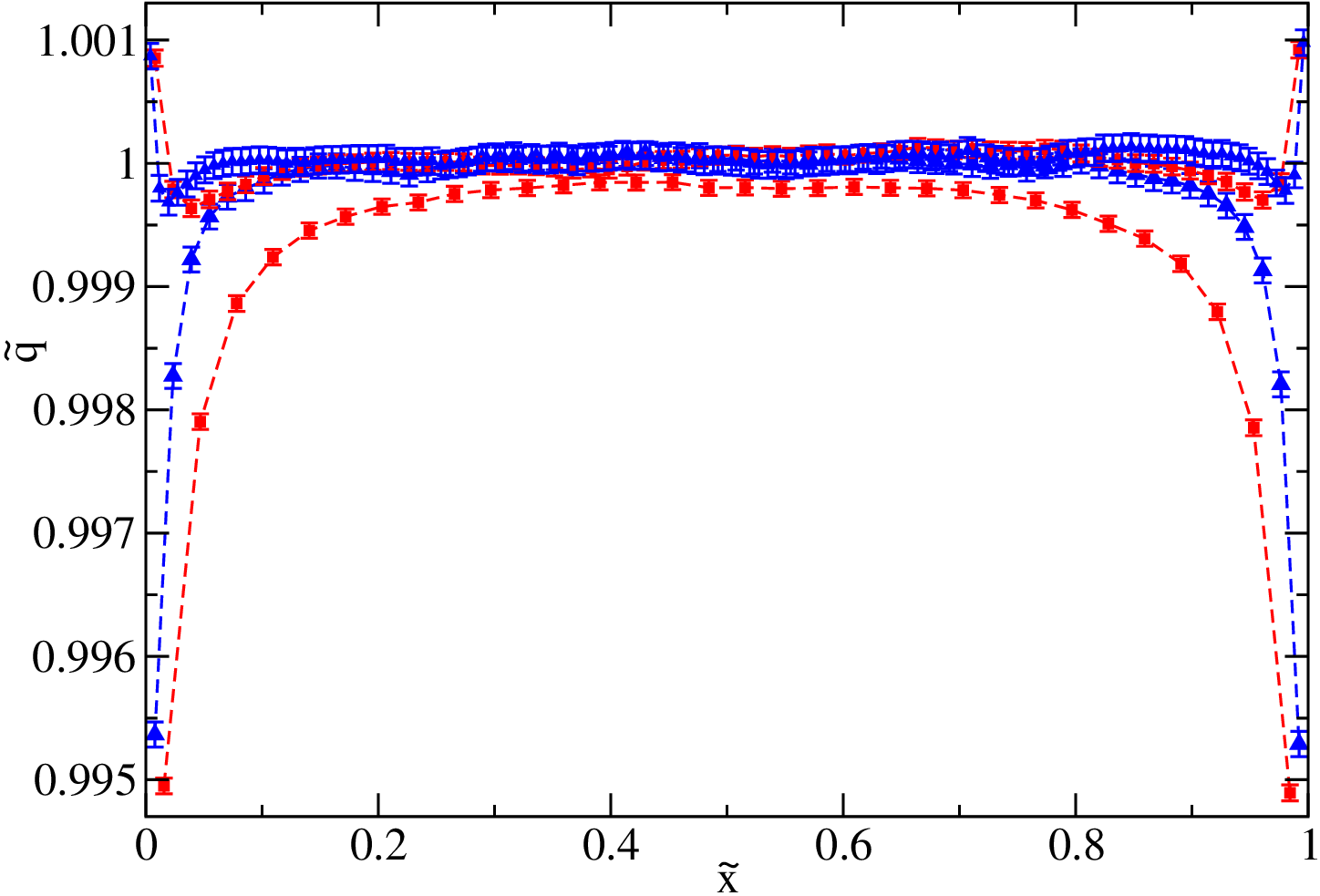}
\caption{\label{magplot2}
Profiles, eq.~(\ref{qdef}), at $c_{1,b}=1.068$, $c_{2,b} = 0.9234$, $\beta=0.387721735$, 
$\beta_{1,2} = 0.185807$ and the lattice sizes $L=64$ and $128$.
We plot $\tilde q=q(x_0)/\bar{q}$ as a function of $\tilde x= x_0/L$ and $ \tilde x= (x_0-L)/L$ for the
fine and the coarse lattice, respectively. The curves for the different lattice sizes can be
identified by the number of data points. Furthermore $\tilde q < 1$ for the fine lattice
and $\tilde q > 1$ for the coarse one. The dashed lines should only guide the eye.
}
\end{figure}

In Fig. \ref{magplot2} we plot $\tilde q=q(x_0)/\bar{q}$, eq.~(\ref{qdef}), 
for $(c_{1,b},c_{2,b})=(1.068,0.9234)$, $\beta=0.387721735$ and $\beta_{1,2} = 0.185807$. 
The improvement compared with $(c_{1,b},c_{2,b})=(1,1)$ potted in Fig. \ref{magplot}
is fairly obvious. Starting from a distance of roughly $10$ lattice spacings from the boundary 
we find that $q(x_0)$ is consistent, at the level of our statistical accuracy, with the result 
obtained for the homogeneous system.
This observation holds for both the fine as well as for the coarse lattice.
We checked also the behavior of $E_0(x_0)$ and $E_{12}(x_0)$, eqs.~(\ref{E0def},\ref{E12def}), 
as a function $x_0$. Here we find that starting from a  distance of roughly $13$ lattice 
spacings from the boundary $E_0(x_0)$ and $E_{12}(x_0)$ are consistent with the results 
obtained for the homogeneous system. 

To be on the safe side, studying properties of the homogeneous system, one should keep at least
a distance of $\approx 20$ lattice spacings from the  boundary between different 
lattice spacings. 

\subsection{Going off-critical}
The coupling $\beta_2$ in the bulk of the coarse lattice as a function of the coupling 
$\beta_1$ of the fine lattice is chosen such that the bulk correlation length is the same  
for both parts of the lattice. Hence in terms of the correlation length $\xi$ in units of the respective
lattice spacing we get
\begin{equation}
 \xi(\beta_1)  = 2 \xi(\beta_2) \;.
\end{equation}
In the neighborhood of the critical point, the correlation length behaves as
\begin{equation}
 \xi(\beta) = \xi_0 |t|^{-\nu} \; (1 + b |t|^{\theta}  + c t  + d |t|^{\theta'} + e |t|^{2 \theta} + ...) \;,
\end{equation}
where $t = \beta_c - \beta$ is our definition of the reduced temperature.
Here we discuss an improved model which is characterized by $b \approx 0$. Hence, in 
the following  we shall neglect the terms $b |t|^{\theta}$ and $e |t|^{2 \theta}$.
Note that $\theta' = \nu \omega'$, where $\omega' = 1.67(11)$, ref. \cite{NewmanRiedel}.
Hence in the numerical analysis of data for the correlation length it is difficult
to disentangle the corrections $c t$ and $d |t|^{\theta'}$.  
After a preliminary study, we realized that for our purpose  
corrections to scaling can be ignored entirely. Hence we get the linear relation
\begin{equation}
\label{fixingb2}
 (\beta_c - \beta_2) =  (\beta_c - \beta_1) \; 2^{1/\nu} \;\;,
\end{equation}
where we take $\nu=0.63002$, ref. \cite{mycritical}.  In order to keep things simple, 
$c_{1,b}$ and $c_{2,b}$ keep their values as determined above at the 
critical point.

It remains to determine $\bar \beta_{1,2}$ as a function of $\beta_1$.  Analogous to
eq.~(\ref{tunebeta12}) we impose
\begin{equation}
\label{tunebeta12off}
 G_{1,2}(L,\beta_1,\bar{\beta}_{1,2}) = \bar{G}(L,\beta_1) := 
 \frac{G_{1,2,hom}(L/2,\beta_2) + G_{1,2,hom}(L,\beta_1)}{2} \;.
\end{equation}
We have simulated $L=24$ and $48$ at one value of $\beta_1$ in the high 
temperature phase and one value of $\beta_1$ in the low temperature phase.
In order to check the dependance of the result on the difference of these two 
values of $\beta_1$, we simulated $L=32$ at four values of $\beta_1$. 
The corresponding results for the slopes $\Delta \bar \beta_{1,2}/\Delta \beta_1$
are given in table \ref{Tslope}. 

\begin{table}
\caption{\sl \label{Tslope}
Results for the slopes $\Delta \bar \beta_{1,2}/\Delta \beta_1$ at  $(c_{1,b},c_{2,b})=(1.068,0.9234)$
}
\begin{center}
\begin{tabular}{clll}
\hline
$L$ &  \mc{1}{c}{$\beta_1$,high} & \mc{1}{c}{$\beta_1$,low} &  \mc{1}{c}{$\Delta \bar \beta_{1,2}/\Delta \beta_1$} \\
\hline
24  & 0.386  & 0.39  & 0.9738(38) \\

32  & 0.386 & 0.3895 & 0.9706(51)  \\
32  & 0.387 & 0.3885 & 0.9774(68) \\

48  & 0.3871& 0.3883 & 0.9663(99)  \\
\hline
\end{tabular}
\end{center}
\end{table}

Within the numerical accuracy we see no non-linearity nor a
dependence on $L$. As our final result we take
\begin{equation}
\frac{\mbox{d} \bar \beta_{1,2}} {\mbox{d} \beta_{1}} = 0.974(10) 
\end{equation}
at the critical point.

\subsection{Summary of our results}
\label{summaryC}
The relation of the bulk coupling on the fine $\beta_1$ and the 
coarse lattice $\beta_2$ is given by eq.~(\ref{fixingb2})
\begin{equation}
 (\beta_c - \beta_2) =  (\beta_c - \beta_1) \; 2^{1/\nu} \;\;,
\end{equation}
where we take $\beta_c=0.387721735$ and $\nu=0.63002$, ref. \cite{mycritical}.
The amplification of the coupling in the layers at the boundary 
of the coarse and the fine lattice are 
\begin{equation}
 c_{1,b} = 1.068 \;\;,\;\;\;\;  c_{2,b} = 0.9234 \;.  
\end{equation}
For these choices  we get for the coupling between the fine and 
the coarse lattice
\begin{equation}
 \beta_{1,2} = 0.185807 + 0.974 \; (\beta_1-\beta_c) \;\;.
\end{equation}
We recommend to use this approach in a range of couplings
give by
\begin{equation}
 0.386 \lessapprox \beta_1  \lessapprox 0.391  \;.
\end{equation}
Note that $\xi_{2nd}(0.386) = 12.5747(24)$. For smaller correlation
length there is little need to use systems with varying lattice
spacing.

\section{Benchmark: Films with strongly symmetry breaking boundary conditions}
We intent to use the results obtained here in the study of the thermodynamic
Casimir force between a sphere and a plane.  To this end, in the neighborhood
of the plane, the lattice spacing $a=1$ is used. Also the sphere should be 
well inside the part of the lattice with lattice spacing $a=1$.  At  increasing 
distances from the plane, larger and larger lattice spacings are used. 
This way, a semi-infinite system could be approximated at relatively low 
computational costs. The systematic error that is introduced by the 
coarsening of the lattice could be controlled by varying the thickness of 
the sectors of a given lattice spacing, in particular the one next to the 
plane.  

Here, as a benchmark and preliminary step, we study the thermodynamic Casimir 
force between two plates.
This geometry is usually called film geometry. In particular we shall simulate
systems with strongly symmetry breaking boundary conditions. This problem 
has been thoroughly studied. For example, in refs. \cite{VaGaMaDi08,MHcorrections} 
Monte Carlo simulations of the Ising model on the simple cubic lattice
were performed. In refs. \cite{MHstrong,MHcorrections,myRecent} the 
improved Blume-Capel model on the simple cubic lattice, the model that we
consider here, has been studied. Hence accurate numbers to compare with are
available. In addition to the  thermodynamic Casimir force we determine 
the magnetization profile. Films with strongly symmetry breaking boundary
were also studied for example by using the extended de Gennes-Fisher local-functional method 
\cite{BoUp98,BoUp08,UpBo13}.

\subsection{Thin films with strongly symmetry breaking boundary conditions}
We study a lattice of the size $(L_0+2) \times L_1  \times L_2$, where 
$L_1=L_2=L$ throughout. In $1$- and $2$-direction periodic boundary conditions 
are imposed. In our convention $x_0 \in \{0,1,...,L_0+1\}$. Strongly symmetry
breaking boundary conditions are implemented by fixing the spins at $x_0=0$ and 
$L_0+1$ to either $s_x=-1$ or $s_x=1$. In the case of $(+,+)$ boundary
conditions, the spins at both boundaries $x_0=0$ and $L_0+1$ are fixed to $s_x=1$,
while for $(+,-)$ boundary conditions, the spins at $x_0=0$ are
fixed to $s_x=1$ and those at $L_0+1$ are fixed to $s_x=-1$.
In order to determine the thermodynamic Casimir force, we simulated two films 
of the thickness $L_{0,1} = L_0+1/2$ and $L_{0,2} = L_0-1/2$.  
In the following $L_{0,2}$ is always a multiple of $4$. In our simulations
we used two different lattice spacings only. In particular, 
we kept the layers $x_0=0$, $1$, ..., $L_{0,2}/4$ and 
$x_0=3 L_{0,2}/4+1$, $3 L_{0,2}/4+2$,  ... $L_{0,i}+1$ as fine lattice. The remaining
ones, in the center of the film, are replaced by a coarse lattice with twice the 
lattice spacing.

The thermodynamic Casimir force per area is defined by
\begin{equation}
\label{definecasimir}
\beta F_{Casimir} = -   \frac{ \partial f_{ex} }{ \partial L_0} \;\;,
\end{equation}
where  $f_{ex} = f_{film} - L_0 f_{bulk}$ is the reduced excess free energy per area of the 
film of the thickness $L_0$ and $f_{film}$ is the reduced free energy per area of the film
and $f_{bulk}$ is the reduced free energy density of the bulk system.  On the lattice 
the partial derivative~(\ref{definecasimir}) is approximated by a finite difference
\begin{equation}
\frac{ \partial f_{ex} }{ \partial L_0} \approx \Delta f_{ex} 
\end{equation}
where 
\begin{equation}
 \Delta f_{ex}(L_0) = f_{ex}(L_0+1/2) - f_{ex}(L_0-1/2) \;.
\end{equation}
Here, following Hucht \cite{Hucht} we compute $\Delta f_{ex}$ by integrating the corresponding 
difference of excess internal energies per area
\begin{equation}
  \Delta f_{ex} = - \int_{\beta_0}^{\beta} \mbox{d} \tilde \beta \Delta E_{ex}
\end{equation}
where
\begin{equation}
\Delta E_{ex}(L_0) = \Delta E(L_0) - E_{bulk}
\end{equation}
where $E_{bulk}$ is the energy density of the bulk system and
\begin{equation}
\Delta E(L_0) = E(L_0+1/2) - E(L_0-1/2) 
\end{equation}
where we define the energy per area of the homogeneous system as
\begin{equation}
\label{energyf}
 E = \frac{1}{L_1 L_2} \sum_{<xy>} \langle s_x s_y \rangle \;.
\end{equation}

Now let us discuss how this approach can be generalized to our system 
with two different lattice spacings.
The partition function takes the form
\begin{equation}
 Z =\sum_{\{S\},\{s\}}  \exp\left( \beta e_0(\{s\})
                  + \sum_i (a_i +b_i \beta) e_i(\{S\},\{s\}) 
                   - D \sum_x s_x^2
                   - D \sum_X S_X^2
 \right)
\end{equation}
where the $\{S\}$ are the spins on the coarse lattice and $\{s\}$
the ones on the fine lattice.
The $e_i$ correspond to sums in eq.~(\ref{mixedaction}).
Taking the derivative of the reduced free energy per area with respect to 
$\beta$ we arrive at
\begin{equation}
\frac{\partial f}{\partial \beta} = -  \frac{1}{L_1 L_2} 
\frac{\partial \ln Z}{\partial \beta} = - \frac{1}{L_1 L_2}
\left[\langle e_0 \rangle + \sum_i b_i \langle e_i \rangle    \right ]
\end{equation}
which means that eq.~(\ref{energyf}) is generalized to
\begin{equation}
 E =  \frac{1}{L_1 L_2} 
\left[\langle e_0 \rangle + \sum_i b_i \langle e_i \rangle \right ] \;\;.
\end{equation}

\subsection{Numerical results}
Here we have repeated a part of the simulations discussed in 
section 5 of ref. \cite{myRecent}.
The parameters of the simulation are very similar 
to those of ref. \cite{myRecent}. In particular we used the exchange 
cluster algorithm and the improved estimator that is associated with 
it, to compute $\Delta E$.  The implementation of the exchange 
cluster algorithm for the system with two different lattice spacings 
and the associated Hamiltonian is straight forward.

Let us first discuss the results obtained for $\Delta E $ for $(+,+)$
boundary conditions. We simulated
$(+,+)$ boundary conditions for $(L_0,L)= (16.5,64)$,  $(32.5,128)$,
and $(64.5,256)$ at about 100 values of $\beta$ in the neighborhood of the
critical point.
\begin{figure}[tp]
\vskip0.5cm
\includegraphics[width=13.5cm]{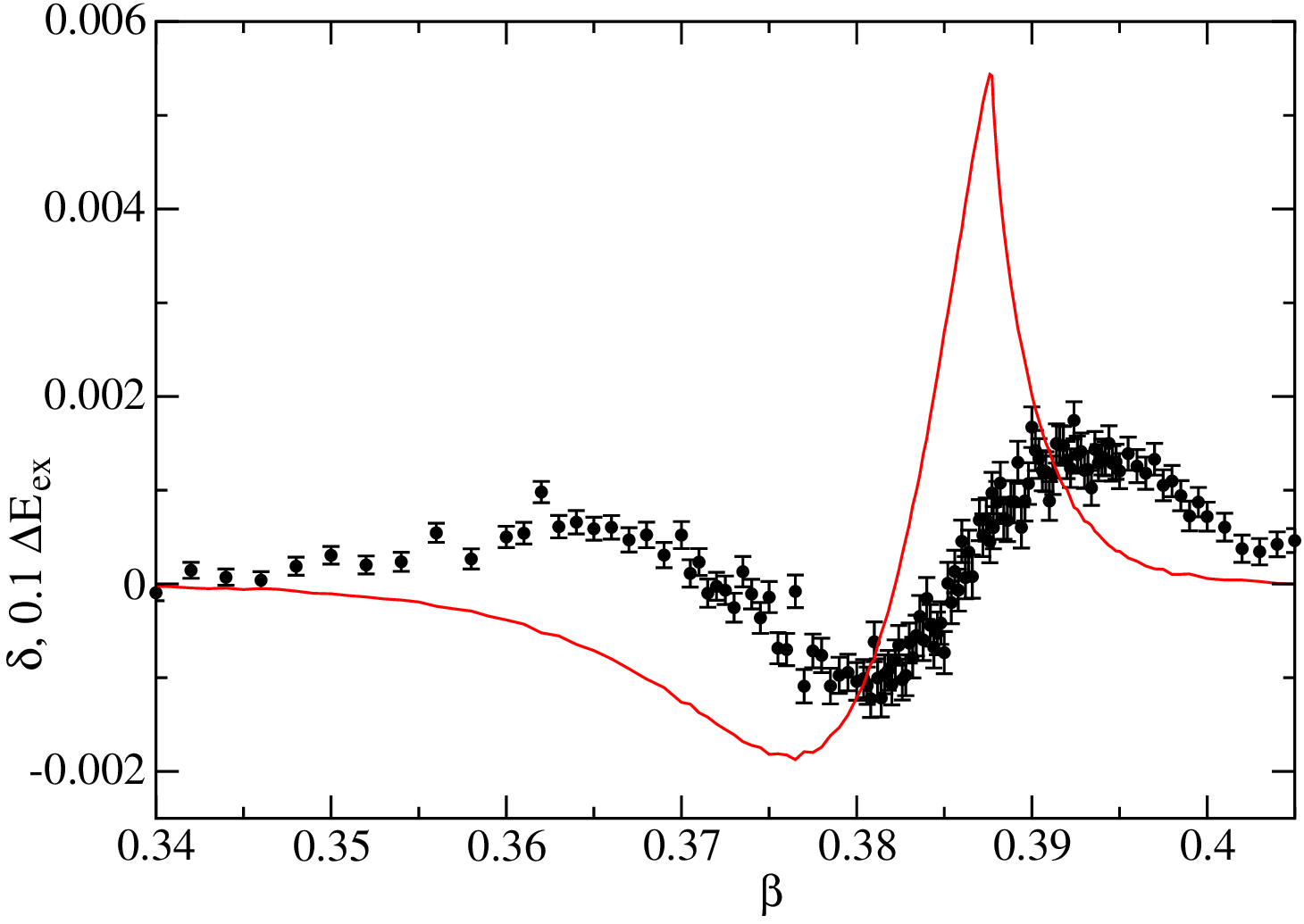} 
\vskip1.0cm
\includegraphics[width=13.5cm]{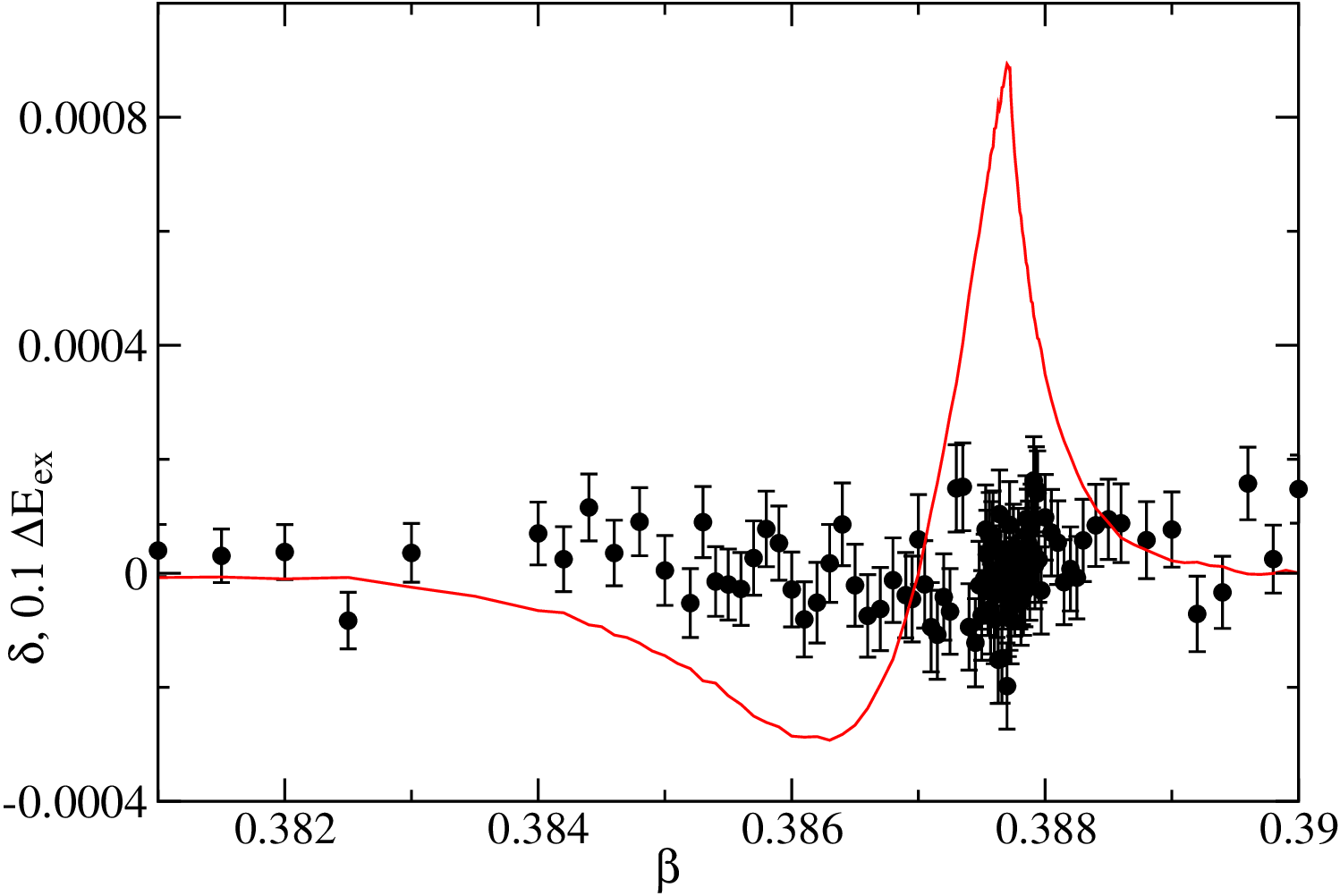}
\caption{\label{diffP16}
We plot $\delta=\Delta E - \Delta E_{hom}$ for $(+,+)$ boundary
conditions as a function of $\beta$. 
In addition we show as solid red line, $0.1 \Delta E_{ex,hom}$.
Results for the lattice size $(L_0,L)=(16.5,64)$ and $(L_0,L)=(64.5,256)$ 
are given in the upper and lower part of the figure, respectively.
}
\end{figure}

In the upper part of Fig. \ref{diffP16} we plot the difference 
$\delta=\Delta E - \Delta E_{hom}$ and $0.1 \Delta E_{ex,hom}$ for 
$L_0 = 16.5$ and $L=64$,
where $hom$ indicates the result for the homogeneous system, taken from 
ref. \cite{myRecent}. We see that $\delta$ is definitely different 
from zero for most of the values of $\beta$ that we have simulated at.
On the other hand, on average $|\delta|$ is much smaller than 
$|\Delta E_{ex,hom}|$. Therefore the result that we get for the 
thermodynamic Casimir force is at least qualitatively correct. Going to 
larger thicknesses the error due to the coarsening in the center of the 
film becomes less and less visible. While for $L_0=32.5$ still
$|\delta| \ne 0$ at the level of our statistical accuracy, for $L_0=64.5$
a deviation of $\Delta E$ from $\Delta E_{hom}$ can hardly be seen.
See the lower part of Fig. \ref{diffP16}.  

\begin{figure}[tp]
\vskip0.5cm
\includegraphics[width=13.5cm]{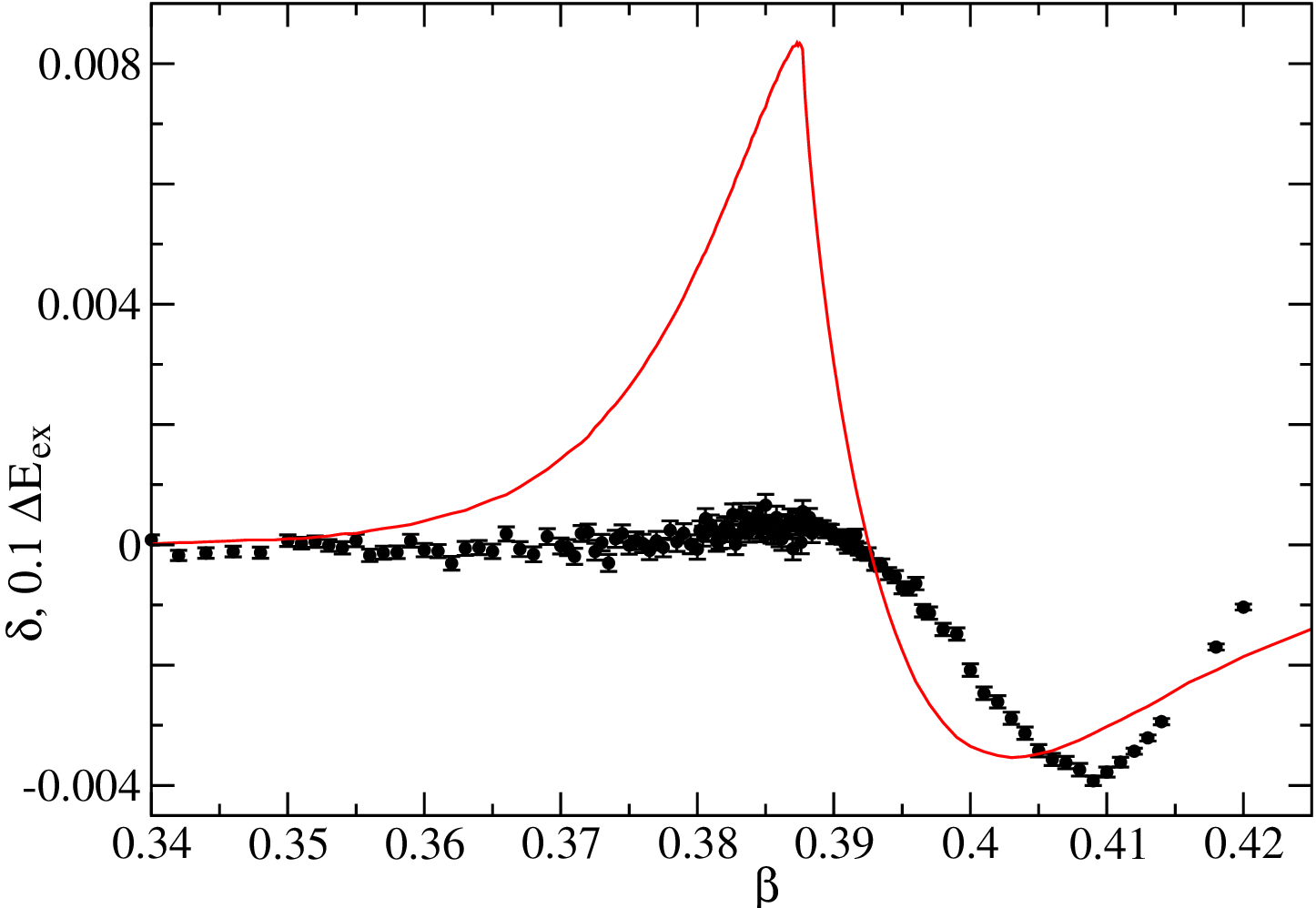}
\vskip1.1cm
\includegraphics[width=13.5cm]{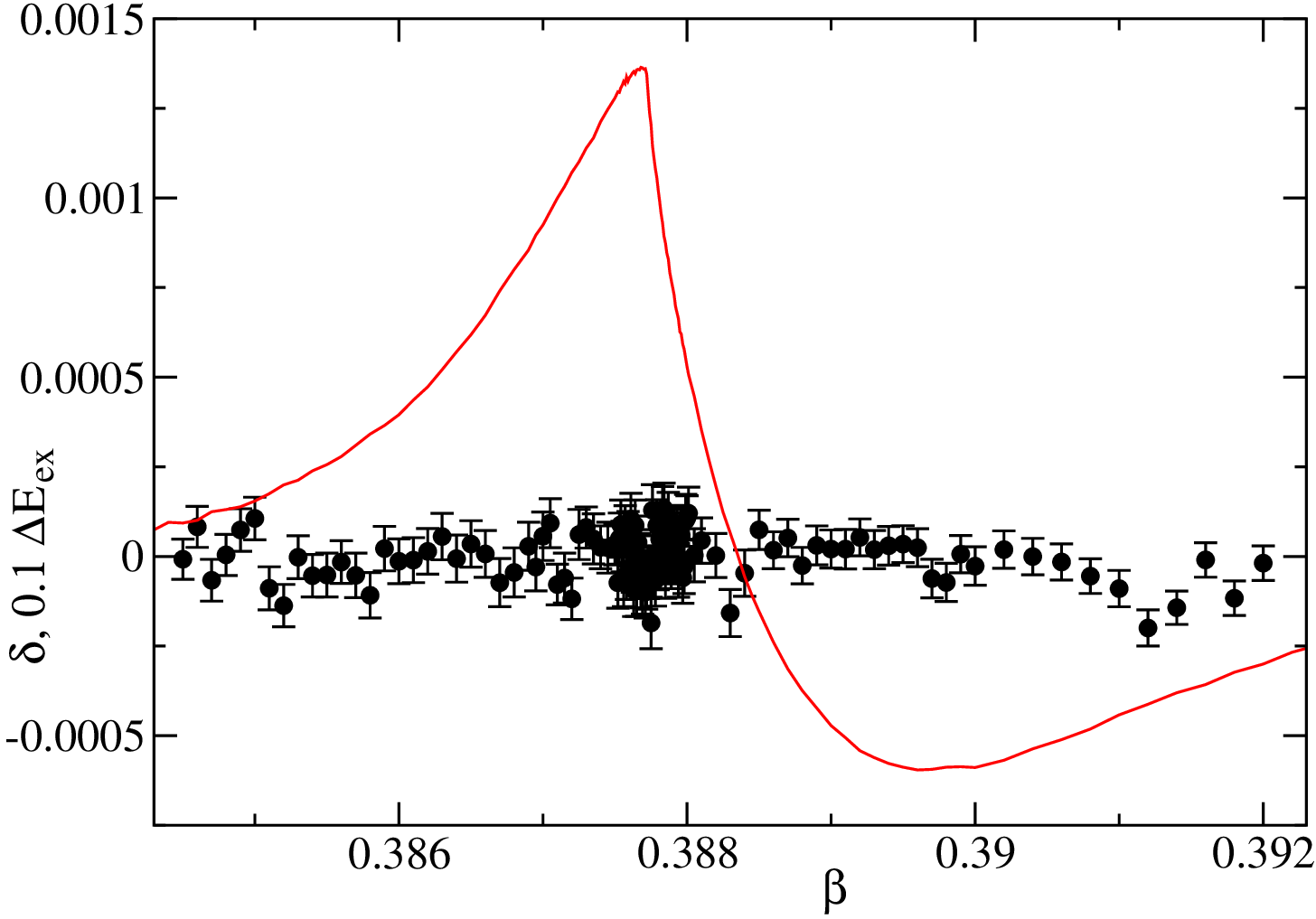}
\caption{\label{diffA16}
Same as Fig. \ref{diffP16} but for $(+,-)$ instead of $(+,+)$ boundary
conditions.
}
\end{figure}

Let us turn to $(+,-)$ boundary conditions. We simulated films of the 
thicknesses $L_0=16.5$, $32.5$ and $64.5$. For each thickness we simulated
at several values of $L$. For example for $L_0=16.5$ we studied
$L=64$, $128$ and $256$, where $L$ is increased with increasing $\beta$.
For a more detailed discussion see ref. \cite{myRecent}. Our numerical
results for $\delta=\Delta E - \Delta E_{hom}$  for $L_0=16.5$ and $64.5$
are shown in Fig. \ref{diffA16}.
For $L_0=16.5$ we see  a strong deviation of $\Delta E$ from the corresponding
result for  homogeneous system in the low temperature phase.
This deviation is only one order of magnitude smaller than the quantity 
$\Delta E_{ex,hom}$ that we like to compute. For the thickness $L_0=32.5$
the systematic error $\delta$ is still clearly visible and 
its functional form is similar to that for $L_0=16.5$. As for 
$(+,+)$ boundary conditions, for $L_0=64.5$ we find that 
$\delta$ is essentially consistent with zero at the level of our 
statistical accuracy.

As an additional  check, we computed for $L_0=64.5$ the integral 
\begin{equation}
 \Delta = \int_{\beta_0}^{\beta} \mbox{d} \tilde \beta \; \delta
\end{equation}
which gives the error of $\Delta f$. 
In the case of $(+,+)$ boundary conditions, we find that $\Delta$ is 
consistent with zero within two standard deviations in the whole
range of $\beta$ that we have studied. The some observation holds
for $(+,-)$ boundary conditions up to $\beta=0.391$. Then $|\delta|$ is
increasing. For  $\beta=0.392$ we see a deviation from zero by a little
less than three standard deviations.

\subsubsection{Magnetization profile} 

The magnetization profile of the homogeneous system is defined by
\begin{equation}
 m(x_0) =  \frac{1}{L_1 L_2}  \left \langle \sum_{x_1,x_2}   s_x \right \rangle \; .
\end{equation}
Note that for $(+,\pm)$ boundary conditions 
\begin{equation}
\label{Zsymm}
m(x_0) = \pm m(L_0+1-x_0) \; .
\end{equation}
For a discussion of the behavior at the critical point see \cite{MHstrong}
and refs. therein. Numerical results are given in section VI. B of 
\cite{MHstrong}, see in particular Fig. 1.  For simplicity we only discuss
the second system that we simulated, since for
$L_{0,2}=16$, $32$ and $64$ in contrast to $L_{0,1}=17$, $33$ and $65$, 
the symmetry~(\ref{Zsymm}) is not broken by the coarsening.

\begin{figure}[tp]
\vskip1.0cm
\includegraphics[width=13.5cm]{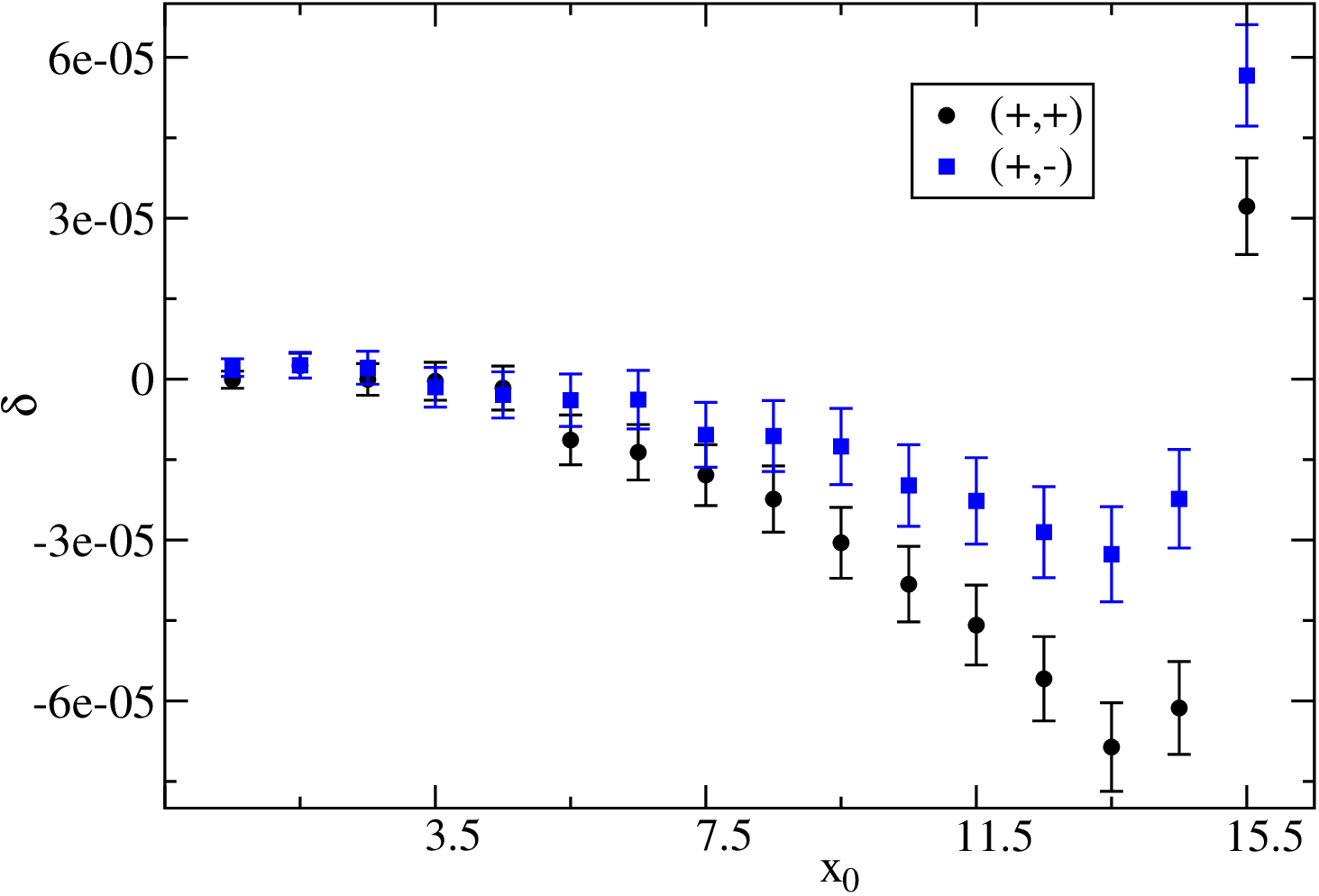}
\caption{\label{diffmag}
We plot the difference $\delta m(x_0)$, eq.~(\ref{diffm}),  between the magnetization 
profiles of the composite and
the homogeneous system for both $(+,+)$ and $(+,-)$ boundary conditions
and the thickness $L_0=64$ at the critical temperature.
We averaged $\delta m(x_0)$ over all values $0.3852 \le \beta \le 0.38792$
we simulated at.
}
\end{figure}

Let us first discuss the results obtained for the fine part of the lattice
with those for the homogeneous system for the same lattice size. 
In Fig. \ref{diffmag} we plot the difference
\begin{equation}
\label{diffm}
 \delta m(x_0) =  m(x_0) - m_{hom}(x_0)
\end{equation}
for $L_0 = 64$ for both $(+,+)$ and
$(+,-)$ boundary conditions. For individual values of $\beta$  we 
can hardly distinguish $\delta m(x_0)$ from zero. Therefore, we have 
averaged $\delta m(x_0)$ over all values $0.3852 \le \beta \le 0.38792$
we simulated at.  
This way a small systematic error is revealed.

We also determined $\delta m$ for the thicknesses $L_0=16$, $32$. Already 
here the systematical error is rather small.

\begin{figure}[tp]
\vskip1.0cm
\includegraphics[width=13.5cm]{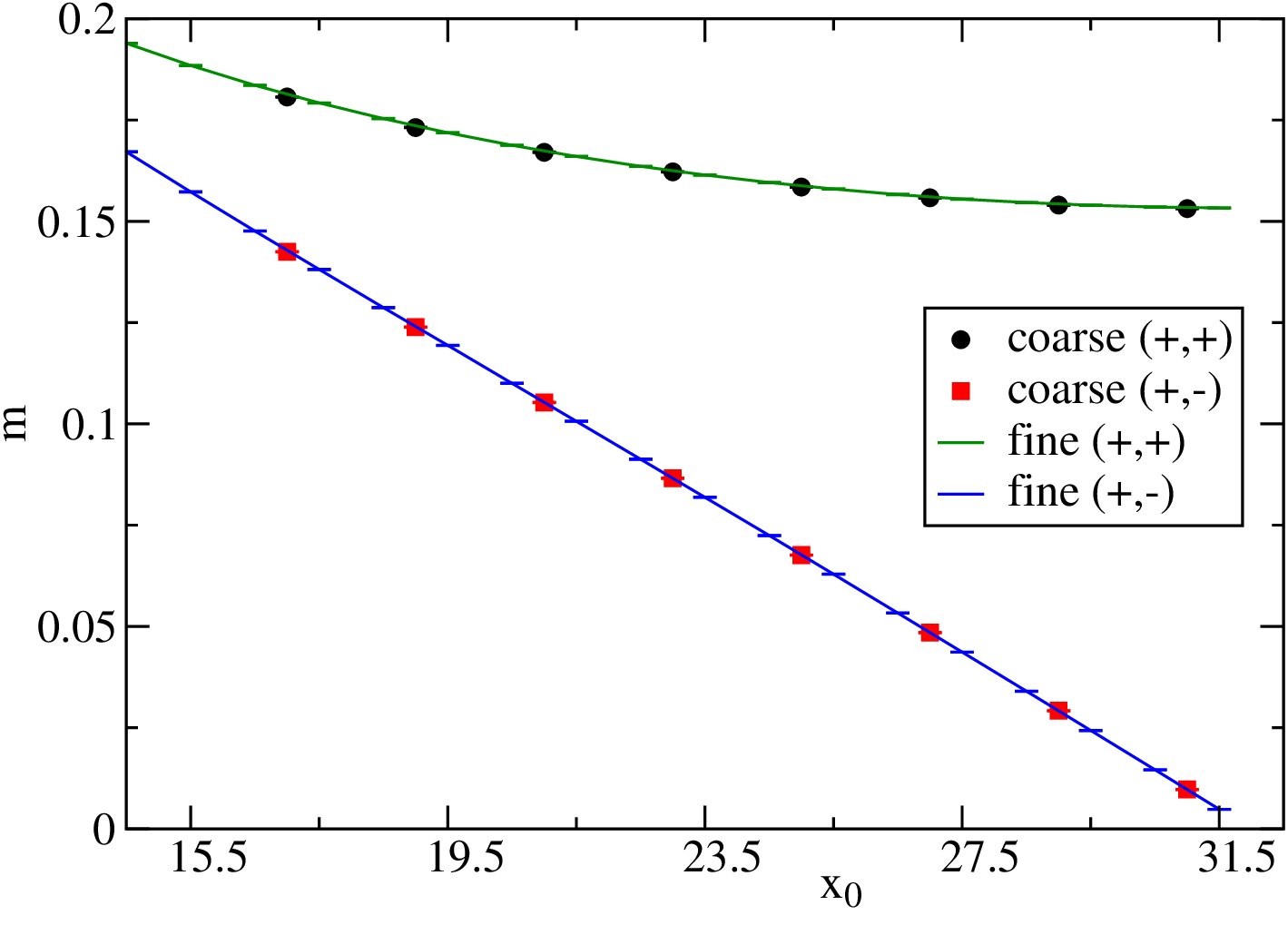}
\caption{\label{magprofC64}
We plot the magnetization profile for both $(+,+)$ and $(+,-)$ boundary conditions
and the thickness $L_0=64$ at the critical temperature. The data points
denoted by ``fine'' are taken from simulations of the homogeneous system. The data
point taken from the coarse part of the composite lattice are rescaled by a factor
of $2^{-(1+\eta)/2}$. In the case of the fine lattice, the solid lines should 
only guide the eye.
}
\end{figure}

For the coarse part of the lattice the comparison is less direct. 
First one has to rescale the magnetization profile by a factor of 
$2^{-(1+\eta)/2}$ to compare it with the corresponding profile
of the homogeneous system. Moreover, the coarse sites are located at
integer values of $x_0$, while in our convention the sites of the fine
lattice assume half integer values of $x_0$. Therefore also an interpolation of 
the profile is needed for the comparison. Here we do not  
delve into this direction but simply plot our data. In Fig. \ref{magprofC64}
we plot the magnetization profile at $\beta=0.387721735$ for 
$(+,+)$ and $(+,-)$ boundary conditions. We give the results obtained 
for the homogeneous system and those for the coarse lattice that we
have rescaled as discussed above.
At least by eye one can only see a deviation in the case of $(+,+)$ 
boundary conditions at $x_0=17$, next to the boundary between  
parts of the lattice with different lattice spacings. The magnetization on the 
coarse lattice seems a bit too small compared with the one of the 
homogeneous lattice.

\section{Conclusions and Outlook}
We aim at studying critical phenomena using lattices with adaptive
lattice spacing. This should be useful in studies of systems 
with nontrivial boundary conditions. 
We are planning to study the thermodynamic Casimir force between 
a planar and a spherical object or between two spherical objects.
In critical phenomena an 
adaptive grid is difficult to implement, since the field and 
the couplings rescale nontrivially under a change of the length scale.
The present work is a first step in this direction. We consider
the universality class of the three-dimensional Ising model. Our 
starting point is the improved Blume-Capel model on the simple
cubic lattice with lattice spacing $a$. The general idea is to replace 
the sites within certain sectors of the lattice by coarse sites
that are separated by a lattice spacing $2 a$. This procedure 
might be iterated to lattice spacings $4 a$, $8 a$, ... . 
Furthermore these sectors are composed of cubes with faces that are 
parallel to the lattice-axis. 
Here we made a first step in this program. In section \ref{themethod} we worked
 out, how two 
half spaces with lattice spacing $a$ and $2 a$, respectively, that 
are separated by a boundary that is perpendicular to one of the lattice
axis, can be coupled consistently. To this end, we used finite
size scaling. The boundary between  lattice spacing $a$ and $2 a$
can be viewed as a defect plane. Following refs. \cite{BuEi81,DiDiEi83},
there is one relevant perturbation with the RG-exponent $y=y_t-1$, 
where $y_t=1/\nu$, associated with it. Hence one would expect that 
it is sufficient to tune the coupling between the two half spaces.
However it turned out that slowly decaying corrections remain.
These can be removed by tuning the couplings within the layers next
to the boundary.
Our final results for the coupling constants are summarized in section
\ref{summaryC}.
In a more general scenario we have also edges and corners. 
These can be regarded as line and point defects, respectively. 
Following the argument given in the appendix of \cite{BuEi81} the 
RG-eigenvalues are $y_l=y_t-2 = -0.41275(25)$ and $y_p=y_t-3 = -1.41275(25)$, 
respectively. Hence both perturbations of the fixed point are irrelevant.
Therefore in principle no additional tuning of couplings at the edges and 
corners is needed. However the absolute value of $y_l$ is rather small
and hence still tuning of a coupling at the edges seems to be 
beneficial. It remains to be worked out which coupling could be tuned
and which observables could be used to tune it.

In order to benchmark our results we simulated films with strongly symmetry 
breaking boundary conditions. We replaced half of the sites, located 
in the center of the film, by coarse sites.  We computed the magnetization
profile and the partial derivative of the internal energy with respect 
to the thickness of the film. This quantity is used to determine the 
thermodynamic Casimir force. For the largest thickness $L_0=64.5 a$ that
we have simulated, at the level of our statistical accuracy, the numerical 
results can hardly be discriminated from those obtained by simulating a 
system with a unique lattice spacing.

\section{Acknowledgement}
This work was supported by the DFG under the grant No HA 3150/3-1.

\end{document}